\RequirePackage{fix-cm}
\documentclass[twocolumn,epjc3]{svjour3}  
\smartqed  % flush right qed marks, e.g. at end of proof
\RequirePackage{graphicx}
\RequirePackage{amssymb}
\journalname{Eur. Phys. J. C}
\begin{document}

\title{Detection  of  precessing circumpulsar  disks%\thanksref{t1}
}


\author{Catia Grimani\thanksref{e1,addr1}}

\thankstext{e1}{e-mail: catia.grimani@uniurb.it}


\institute{DiSBeF, Universtit\`a degli Studi di Urbino ``Carlo Bo'', Urbino (PU) and Istituto Nazionale di Fisica Nucleare, Florence, Italy \label{addr1}}

\date{Received: date / Accepted: date}

\maketitle

\begin{abstract}
%Pulsar initial  parameters affect   formation of  supernova fallback disks  near the light cylinder 
%and magnetospheric particle acceleration.
%We show  that multimessenger clues on these key parameters are compatible with the possibility  that
%a large fraction of  pulsars are  surrounded by  fallback disks.
Experimental evidences indicate that formations of disks and planetary systems around pulsars are allowed.
Unfortunately, direct detections through electromagnetic observations appear to be  quite rare.
In the case of PSR 1931+24, the hypothesis of a rigid precessing disk penetrating the pulsar light cylinder 
is found consistent with radio transient observations from this star. Disk self-occultation and 
precession may limit electromagnetic observations. 
Conversely, we show here that  gravitational waves  generated by  disk precessing near the light cylinder 
of young and middle aged pulsars 
 would be detected by  future  space interferometers  with sensitivities like those 
expected for DECIGO (DECI-hertz Interferometer Gravitational 
Wave Observatory)  and BBO (Big Bang Observer). The 
characteristics of  circumpulsar detectable precessing disks are estimated  as a function of distance from 
the Solar System. Speculations on upper limits to detection rates are presented.
\end{abstract}

\section{Introduction}
\label{intro}
%Pulsars are  considered among the most fascinating astrophysical sites.
%They were proposed, for instance, as sources of high-energy cosmic rays  
%(see, for example, \cite{giller,feng})  and  the
%pulsar magnetosphere   would be a  plausible source of cosmic-ray e$^+$  (see \cite{gri07aa} and references therein). 
%A good knowledge of  pulsar physics and of the pulsar environment would provide precious clues   in  many  fields in Astrophysics.
The observations of coplanar planets surrounding the millisecond pulsar PSR 1257+12  \cite{frail}
and  of a disk
around the anomalous X-ray pulsar (AXP) 4U 0142+61  \cite{wang}
seemed to suggest   that
circumpulsar disk formation  plays an important role in  pulsar spin down \cite{menou,grim09}
and particle  acceleration quenching in the pulsar magnetosphere \cite{grim13}.                                                                                               %Pulsars
%might also  accelerate  high-energy cosmic rays as it was pointed out in \cite{giller,feng} for instance, and  the
%pulsar magnetosphere   would be a  plausible source of cosmic-ray e$^+$  (see \cite{gri07aa} and references therein).

 Jiang and Li \cite{jili} carried out a Monte Carlo simulation of the pulsar evolution by assuming that all
young pulsars could be  surrounded by disks. The comparison of the simulation results with  observations allowed
the authors to estimate
the effects of the propeller torque by constraining the parameters of the new-born pulsars in intervals of reasonable
values.

The presence of circumpulsar disks was also invoked  in the literature to explain observations from  soft gamma repeaters (SGR)
\cite{tong},  AXPs  \cite{chatte},
rotating radio transients  \cite{li2006} and central compact objects \cite{erdeve}.
%The interaction between disk and and pulsar magnetosphere can influence the evolution of the star through the                                                  
%propeller effect.                                                                                                                                              
Grimani  has also shown  \cite{grim09} that  the presence of disks near the light cylinder of
 a large sample of pulsars appears
 compatible with both observed  braking indices and the  e$^+$ production in the galactic pulsar magnetosphere  contributing
to near-Earth positron observations \cite{adri2010,ams13}.

In case of disk formation from supernova fallback material, it was found (\cite{cuha} and references therein) 
that after the supernova explosion
$\le$ 0.1 M$_{\odot}$ of material may remain with typical  angular momentum 
 of 10$^{49}$ ergs s. Material with this  momentum value  would naturally form Keplerian disks
near the light cylinder of young pulsars. However, Perna et al. \cite{perna14} have recently demonstrated that disk formation
around neutron stars occurs only in case of minor magnetic coupling among star layers  during the evolution  and
if the explosion geometry leaving the neutron star behind presents peculiar characteristics.
In addition, small pulsar initial periods  would prevent the formation of disks near the pulsar light cylinder. 
In  \cite{eksi}  it is shown that fallback disk build up would not be allowed near the light cylinder of
pulsars with periods smaller than approximately 40 ms. 
%Because of these conflicting clues arising from models and observations,  upper limits only can be reasonably set to the pulsar-disk system 
%density in our Galaxy. 
% interaction between disk and  pulsar magnetosphere that would influence the evolution of the star through the
%propeller effect.
%The presence of circumpulsar disks was also invoked  to explain observations from  soft gamma repeaters (SGR)  
%\cite{tong},  AXPs  \cite{chatte},
%rotating radio transients  \cite{li2006} and central compact objects \cite{erdeve}. 
%The interaction between disk and and pulsar magnetosphere can influence the evolution of the star through the 
%propeller effect.
%Grimani  has shown previously \cite{grim09} that, in principle,   the presence of disks near the light cylinder of
% a large sample of pulsars appears
% compatible with both observed  braking indices and the  e$^+$ production in the galactic pulsar magnetosphere  contributing
%to near-Earth positron observations \cite{adri2010}.
%Analogous findings were obtained by Jiang and Li \cite{jili} carried out a Monte Carlo simulation of the pulsar evolution by assuming the
%hypothesis that all
%young pulsars could be surrounded by disks. The comparison of the simulation results with  observations allowed 
%the authors to estimate 
%the effects of the propeller torque and to constrain the parameters of the new-born pulsars in intervals of reasonable 
%values.
From the experimental point of view, it is difficult to prove unequivocally  through electromagnetic  observations  that a large sample of pulsars is surrounded
by disks \cite{missdisk}.
Moreover,                                                                                                     
disk self-occultation may occur in case the outer part                                                                                   
of the disks blocks the electromagnetic emission from the inner part. Disk precession might further                                       
reduce the detection probability.

The mentioned conflicting clues arising from models and electromagnetic observations do not allow us to say any final word about  
the presence of disks around a large sample of pulsars in the Galaxy.  However,
in \cite{grim13} we found that gravitational wave (GW) emission from a  precessing disk
around the nearby pulsar B0656+14 could have been observed by the future 
DECI-hertz Interferometer Gravitational
Wave Observatory (DECIGO)
 \cite{decigo1} and Big Bang Observer (BBO) \cite{bbo} devoted to  gravitational wave  
detection  in space near one hertz.

We extend here this approach  to  other  pulsar-disk systems to study their detectability and characteristics 
in the Sun environment.

This manuscript is organized as follows:
section 2 describes  briefly the processes leading to the formation of circumpulsar disks. In section
 3   we find  consistent, independent results  indicating that  pulsar birth  period
average values are  of hundreds of milliseconds
suggesting that disk formation near the pulsar light cylinder is feasible.
% Despite that, the conflicting results among different
%modelizations of long-lived pulsar-disk systems, detection rates would be difficult to 
%estimate beyond rough upper limits, this method would greatly improve the search of circumpulsar 
%disks with electromagnetic observations.  
%suggesting that disk formation near the pulsar light cylinder is, in general, allowed. 
In sections 4 and 5  the  characteristics of  precessing disks around
pulsar light cyclinders are presented. In section 6 we illustrate the advantages of the detection of gravitational
waves generated by precessing disks  near the light cylinder of young and middle aged pulsars  with respect to electromagnetic observations. 

%In case the  presence of precessing disks near the light cylinder of pulsars close to the Solar System 
%will be confirmed,  their  role in affecting  the pair production in the pulsar
%magnetosphere should be properly taken into account.

\section{Origin and effects of circumpulsar disks}

Two possible origins were proposed for circumpulsar  disk formation: supernova fallback
matter and tidal disruption  \cite{cuha}.
In the first case, part of the outer shell of the supernova
fails to escape the gravitational potential of the neutron star.
A disk can form if the fallback material
has enough angular momentum 
or
 when the total pulsar energy output equals the gravitational potential energy
gained by the infalling material forced into rotation by the  electromagnetic interaction with the
pulsar magnetosphere  \cite{michel}.

Metal rich circumpulsar disks are expected to form from supernova fallback material.
Conversely,  disks should consist of light elements in case the circumpulsar material originates from a
companion star tidally disrupted by the pulsar. The first scenario is conceivable with  isolated
pulsars, while the second may likely occur in old, millisecond pulsars.

In the paper by  Menou, Perna and Hernquist \cite{menou} it is found that the characteristics
of some young radio pulsars, including  Crab,  are consistent with the hypothesis
that they are surrounded by fallback disks.
We have also shown  \cite{grim13} that in the case of  PSR B0656+14, a precessing disk \cite{perna} penetrating
the magnetosphere  and quenching the particle production
 may explain the weak pulsed gamma-ray observations from this pulsar \cite{abdo}.
If this is the case,
the role of nearby pulsars
  in contributing to near-Earth positron observations
 at energies larger than a few tens of GeV \cite{bush,hooper,piem} should be reconsidered accordingly.

\section{New-born pulsar initial parameters}

It is commonly assumed that 10$^9$ neutron stars  belong to  our galaxy.
Isolated neutron stars are generally associated with supernova events. 
The  distribution of massive, short lived stars in the Galaxy spiral arms
constrains  the  pulsar spatial  distribution  (see for example \cite{faucher})
and
the supernova explosion rate in the Milky Way of one every 50 years \cite{diehl}
sets an upper limit to the pulsar birthrate (PB). 
The Fermi/LAT collaboration has  found a PB of one pulsar 
born every 59 years on the basis of  a Monte Carlo simulation    
normalized to young $\gamma$-ray pulsar observations  \cite{wr11}.
 Star break up sets the lower limit to the pulsar rotation periods to
about 0.6 ms, however
sub-millisecond pulsars have never been observed \cite{du}.
Maciesiak, Gil and Ribeiro \cite{mac}
report that approximately 1830 pulsars are known  with periods ranging from 1.4 ms to 8.5 s.

The simulation of the active pulsar sample in the Milky Way by  Faucher-Gigu\`ere and Kaspi \cite{faucher}
indicates  pulsar initial periods  of
300$\pm$150 ms.
Experimental clues on  initial periods
are  obtained for those pulsars
with known  ages and braking indices.  For  the pulsars associated with  supernovae G263.9-3.3, SNR 0540-69.3, G11.2-0.3, 
G320.4-1.2 birth periods of
 52 ms, 63 ms, 63 ms and 39 ms were found, respectively (\cite{vande} and references therein).
By using the same method,  for Crab it was found 19 ms while 16 ms and 62 ms were estimated for the pulsars
PSR J0537-6910 and PSR J1811-1925 \cite{faucher}.
The accuracy of this approach may be limited by the assumption that braking indices are constant over the pulsar
age. This is most probably untrue since all known braking indices are smaller than 3,
indicating that pulsars lose energy via processes different from the electromagnetic ones \cite{shap}.
Gravitational wave energy losses \cite{shap} and interaction with surrounding disks \cite{menou,grim09,alpar}
may play some role in pulsar spin down. 
%The resultanting uncertainty on the pulsar 
%initial periods obtained with 
%%%%PB periods
%%%% can be also estimated on the basis of the off-center dipole emission
%%%%in which pulsar proper motion is correlated to initial periods. It has to be
%%%%pointed out that since the orbital velocity of the progenitor and asymmetric kick in
%%%%the supernova explosion may affect the observed pulsar velocity,
%%%%the pulsar birth periods thus obtained must be considered lower limits. By using this approach,
%%%%Huang and Wu  \cite{huang}  find for normal pulsars initial periods of a few milliseconds.
%%%%that for the above considerations do not appear inconsistent               %%%%%with the results given above.                                             
To this purpose, Van der Swaluw and Wu \cite{vande}  considered pulsars residing in composite supernova remnants
consisting of  plerionic and  shell-type components to estimate the pulsar birth  periods.
A sample of 13 composite supernova remnants was studied.
These authors found
that pulsar initial periods vary between 37 and 484 ms. Their findings for the pulsars associated with the
supernovae G263.9-3.3, SNR 0540-69.3, G11.2-0.3, G320.4-1.2 appear consistent with those
inferred from braking indices and consistent with the other results reported above.
We conclude that the uncertainty on the assumption of constant braking indices over the 
pulsar lifetime is negligible.

PBs
 can be also estimated on the basis of the off-center dipole emission
in which pulsar proper motion is correlated to initial periods. It has to be
pointed out that since the orbital velocity of the progenitor and asymmetric kick in
the supernova explosion may affect the observed pulsar velocity,
the pulsar birth periods thus obtained must be considered lower limits. By following this approach,
Huang and Wu  \cite{huang}  estimate normal pulsar initial periods of a few milliseconds.
Vranesevic et al. \cite{vea04}  find that up to 40\% of pulsars are born with periods
ranging between 0.1 and 0.5 s. Similar results were obtained by  Lorimer et al. in \cite{lori06}
and by Vivekahand and Narayan \cite{vn81}.

Theoretical models of cosmic-ray positron production in nearby pulsar magnetosphere \cite{bush,hooper}  lead to a good
agreement with PAMELA \cite{adri2010} data by assuming for these pulsars initial periods of  40 ms or 60 ms.
A pulsar origin for the e$^+$ excess in cosmic rays was also considered in \cite{grim1} with analogous findings.
                                                                                                                                         
Bednarek and Bartosik \cite{bb04,bb05} have shown that their model for high-energy cosmic rays accelerated by pulsars
appears consistent with composition and spectra observations
 between a few$\times$$10^{15}$ eV and a few$\times$$10^{18}$ eV by assuming for pulsar birth periods an
 average value of 400 ms. Giller and Lipski \cite{giller}
and, recently,  Feng, Kotera and Olinto \cite{feng} found consistent results. 

In conclusion,
the majority of observations and models   summarized in Table 1
indicate  that pulsar initial periods
are  of the order of magnitude of tens of milliseconds and above.
Therefore,  in principle,
circumpulsar  disk formation near the light cylinder of young pulsars is allowed  \cite{eksi}.
%In section x 
%we use this result along with those reported in Jinag and Li to 
%to estimate the expected minimum and maximum numbers of pulsar-disk
%systems in the solar environment.

\begin{table}
\caption{\label{table1} Estimates of pulsar initial periods. Table reference legend: VN81 \cite{vn81}; VEA04 \cite{vea04}; FK06 \cite{faucher}.}
%\begin{indented}
%\lineup
\begin{tabular}{lll}%{@{}*{3}{l}}
\hline\noalign{\smallskip}
Author & Pulsar initial &Note\\
&periods (ms) &\\
\noalign{\smallskip}\hline\noalign{\smallskip}
%Pulsar           &  n  \\                                                                                                           
VN81&$>$500&\\
VEA04&$>$ 400&for 40\% of the population\\
FK06       & 300 $\pm$ 150 &    \\
%       B0540-69     & 2.140(9) &     \\
%B0833-45     & 1.4(2)    &   \\
%                                                                                                                                     
% &\0 2013 &\0\0\0\0\0 67 &\0\0\0\0\0 149 & \\                                                                                       
% &\0 2014 &\0\0\0\0\0 54 &\0\0\0\0\0 122 & \\                                                                                       
% &\0 2015 &\0\0\0\0\0 44 &\0\0\0\0\0\0 88 & \\                                                                                      
\noalign{\smallskip}\hline
\end{tabular}
%\end{indented}
\end{table}

\section{Characteristics of disks around isolated young and middle aged pulsars}

\subsection{Supernova mass fallback rate and circumpulsar disk masses}

%The characteristics of circumpulsar disks
%are expected to change with time.
%In particular, as it was recalled above,
%disks surrounding pulsars with ages larger than 10$^5$ years are
%considered to be passive.
%For disks formed
%by supernova infall mass rate.  
%Cannizzo, Lee and Goodman \cite{cann} suggest that the infall mass rate presents a transient
% phase  during which the accretion is nearly constant $\dot{M}${\it (t)}$\simeq${\it M(t}$_o${\it )}
%and after that $\dot{M}${\it (t)} declines in a power-law:

%\begin{equation}
% \dot{M}(t)\ =\ M(t_o)\  \left(\frac{t}{t_o}\right)^{-\alpha}
%\end{equation}

%with t$_o$ $\simeq$ 300 s, $\alpha$=1.25 and  $\dot{M}$(t$_o$)=6.9 $\times$ 10$^{20}$ g s$^{-1}$ - 1.1 $\times$ 10$^{29}$
%g s$^{-1}$. In \cite{mars} the infall matter rate is assumed constant and typically of 10$^{16}$ g s$^{-1}$ - 10$^{17}$
%g s$^{-1}$ as it was found  by  Menou, Perna and Hernquist  for the Crab pulsar \cite{menou}.
%Both modelizations lead to the plausible result that  disk  masses  range in the interval
%between 10$^{-5}$ M$_{\odot}$ - 10$^{-2}$ M$_{\odot}$ (10$^{25}$ kg - 10$^{28}$ kg) as reported in  \cite{grim13} and references therein.

In section 2 we have recalled that circumpulsar disks form if the in-falling matter has an angular momentum or in the case
 the total pulsar energy output, L,  matches  the gravitational potential energy
gained by the in-falling matter:

\begin{equation}
 \dot{M} \frac{G M_o}{r_{lc}}=L
\end{equation}

where G is the gravitational constant, $\dot{M}$ is the in-falling matter rate,  r$_{lc}$=c/$\Omega$ represents the pulsar light cylinder radius with c the speed of light, 
$\Omega$=(2$\pi$)/P, P is  the pulsar period and $M_o$=  1.4 M$_{\odot}$ is the  pulsar mass.

%In the following, we  consider the consequences of  disk formation  near the
%light cylinder of young and middle aged pulsars.
The characteristics of circumpulsar   disks are expected to change with time.
In particular disks surrounding pulsars with ages larger than 10$^5$ years are supposed to be neutral
\cite{cordes,cann}.
For disks formed
by supernova fallback matter,                                                                                                                            
Cannizzo, Lee and Goodman \cite{cann} suggest that the in-falling matter rate presents a transient                                                           
phase  during which the accretion is nearly constant with time t ($\dot{M}${\it (t)}$\simeq${\it M(t}$_o${\it )})                                                         
and after that declines in a power-law:                                                                                                                        

\begin{equation}                                                                                                                                              
\dot{M}(t)\ =\ M(t_o)\  \left(\frac{t}{t_o}\right)^{-\alpha}                                                                                                  
\end{equation}

with t$_o$ $\simeq$ 300 s, $\alpha$=1.25 and  $\dot{M}$(t$_o$)=6.9 $\times$ 10$^{20}$ g s$^{-1}$ - 1.1$\times$10$^{29}$                                                         
g s$^{-1}$. On the basis of this modelization, after 5$\times$10$^5$ years, for instance, the matter fallback rate is  3.68$\times$10$^{6}$
g s$^{-1}$. In \cite{mars} the infall matter rate is assumed constant and typically of 10$^{16}$ g s$^{-1}$ - 10$^{17}$                                                           
g s$^{-1}$ as  found  by  Menou, Perna and Hernquist  for the Crab pulsar \cite{menou}.                                                                                     
Both modelizations lead to the plausible result that  disk  masses  range in the interval                                                                                         
between 10$^{-5}$ M$_{\odot}$ - 10$^{-2}$ M$_{\odot}$ (10$^{25}$ kg - 10$^{28}$ kg) as reported in  \cite{grim13} and references therein. 

\begin{figure}[!t]
  \vspace{5mm}
  \centering
  \includegraphics[width=3.5in]{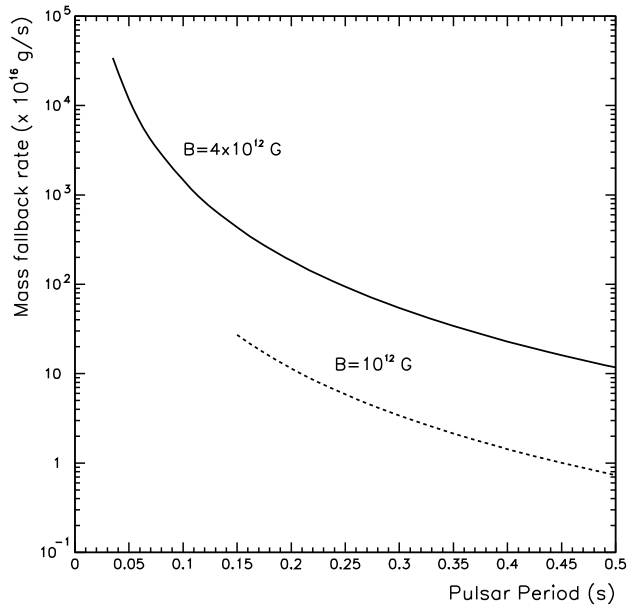}
  \caption{Mass fallback rate allowing for circumpulsar disk building up versus pulsar period. Different
values for the pulsar surface magnetic field (B) were considered for young (4$\times10^{12}$ G; continuous line) 
and middle aged pulsars ($10^{12}$ G; dashed line).}
  \label{fig2}
 \end{figure}

The mass fallback rate allowing for disk formation
can be determined from  equation 1 as a function of the pulsar period \cite{michel}.
Results are reported in figure \ref{fig2}.
Lower and upper limits to
the periods of young and middle aged pulsars   of 40 ms and 500 ms, respectively,  are used as case studies.
Slightly smaller  magnetic fields were considered for middle aged pulsars with respect to younger ones
 (see  \cite{gri07aa} for details).

In figure \ref{fig2} it can be observed that large mass fallback  rates  are needed
for
disk formation  around young pulsars
(as suggested by  Cannizzo, Lee and Goodman \cite{cann}),  while fallback matter rates of $10^{16}$ g s$^{-1}$ - $10^{17}$ g s$^{-1}$
 would be sufficient for disk building up around young and middle aged pulsars with periods of hundreds of milliseconds.

\subsection{Circumpulsar disk dimensions}

The inner radius of disks around young pulsars is set equal to the size of the light cylinder 
by several authors (see for instance \cite{menou})
due to the propeller effect of matter penetrating the light cylinder. We make the same
assumption here by also taking into account the result of the work by Currie and Hansen \cite{cuha}
for the formation of Keplerian disks near the pulsar light cylinder. The outer radius of the disk is 
set    twice the inner radius. This choice follows from similar characteristics  
 of the disk observed around the AXP 4U 0142+61 \cite{wang}.
%In this case, a fraction of fallback material would belong to each ring.
%
%In figure \ref{fig3} slightly smaller  magnetic fields were considered for middle aged pulsars with respect to young ones
%   \cite{gri07aa}.
%

%A precessing disk formed from fallback or interstellar matter near the
%light cylinder of the 1.6 million year old  PSR 1931+24 was invoked
%to explain the dispersed radio emission from this pulsar \cite{xdli}.
%In Grimani, 2011 we have considered an analogous  possibility
%for the pulsar B0656+14 where a precessing disk could  quench
%radio and pair production in the pulsar magnetosphere.

%The pulsar electromagnetic emission depends on wether precessing disks penetrate the light cylinder
%quenching the production and acceleration of particles in the outer gap.

\section{Circumpulsar disk precession}

Disk free precession may result from the misalignment between the
angular momentum of the  disk and the neutron star that most likely receives a kick
during the supernova explosion.
%The disk precession may also be induced by the radiation or magnetic torques
%generated by the neutron star (\cite{xdli} and references therein).
The  dispersed radio emission from the 1.6 million year old  pulsar PSR 1931+24 
is compatible with the presence of a a precessing disk, formed from supernova fallback matter 
or interstellar matter, entering the
light cylinder periodically 
%%%%of the 1.6 million year old  pulsar PSR 1931+24 was invoked
%%%%to explain the observed dispersed radio emission from this star 
\cite{li2006}.
This observation seems to suggest that rigid-body disk precession 
around magnetized neutron stars can be considered  for disks lying outside the 
light cylinder as it was suggested by Pfeiffer and Lai
\cite{pfe} for disks  penetrating  the light cylinder.
The age of PSR 1931+24 seems also to indicate that pulsar-disk systems have long lifetimes.
In Grimani \cite{grim13} we have considered an analogous  possibility
for the pulsar B0656+14 where a precessing disk could  quench
radio and pair production in the pulsar magnetosphere.
%Rigid-body disk precession around magnetized neutron stars was considered by Pfeiffer and Lai 
%\cite{pfe}, for instance.
Precessing disks were  also observed in X-ray binaries and active galactic nuclei
(e.g. \cite{ogdu} and references therein).
%The pulsar electromagnetic emission depends on wether precessing disks penetrate the light cylinder
%quenching the production and acceleration of particles in the outer gap.
%Both mechanisms illustrated above would lead to naturally form disks around the light cylinder of young pulsars,
%but only  disk free precession may allow for the disk characteristics to remain basically unchanged over 
%the entire lifetime of the disk precession as the case of the pulsar PSR 1931+24 seem to suggest.
The Keplerian frequency associated to circumpulsar precessing disks is that corresponding to the disk inner radius \cite{grim09,hake}. 
Pulsars with periods in the interval 40 ms - 500
 ms present light cylinders ranging between  1.9$\times$$10^3$ km and 2.4$\times$$10^4$ km.
Disk dimensions would be equal to 1.9$\times$$10^3$ km - 3.8$\times$$10^3$ km and 
2.4$\times$$10^4$ km - 4.8$\times$$10^4$ km for young
and middle aged pulsars, respectively.
%The reason of having chosen the disk form similar to that of an annulus derives from the 
%characteriscs of protoplanetary disks before planetary formation that was found to occur 
%around the pulsar PSRxxxxx.
%In this case, the whole amount of fallback material would be separated among the 
% various disks. 

%In the next section we discuss the detectability of circumpulsar precessing disks through 
%electromagnetic and gravitational wave observations.

\section{Detectability of  circumpulsar disks}

\subsection{Infrared/optical/X-ray  surveys for circumpulsar disk detection}

After the discovery of the planetary system around the millisecond pulsar PSR 1257+12, searches for circumstellar
matter around neutron stars were conducted around objects of different characteristics and ages with
 near, mid, far-infrared and optical observations. The X-ray
emission is found to provide  precious clues about the origin of the infrared emission from the disk.
In 1993
Van Buren and Terebey \cite{vanbu} searched the IRAS database for far-infrared emission from the locations
of 478 known pulsars. The flux density sensitivity was larger than 500 mJy and the result was 
that none of the stars showed the presence of disks. A following more sensitive search for 10 $\mu$m 
emission from PSR 1257+12 indicated
a flux upper limit of 7$\pm$11 mJy. 
Kock-Miramond et al. \cite{kock} searched for evidence of emission at 15 $\mu$m from six nearby
pulsars, both isolated and in binary systems,  up to a distance of
1 kpc. No emission was detected. For the nearest pulsar J0108-1431 the 3$\sigma$ upper limits on the flux density was
about 66 mJy at 15 $\mu$m and 22.5 mJy at 90  $\mu$m.  The authors of this work set upper limits to the mass of circumpulsar dust in a range between less than 10$^{27}$ kg
and less than 10$^{23}$ kg for pulsar distances up to 1 kpc.
However,  these observations were 
mainly sensitive to dust with T $\gtrsim$ 300 K.
\begin{figure}[!t]
  \vspace{5mm}
  \centering
  \includegraphics[width=3.5in]{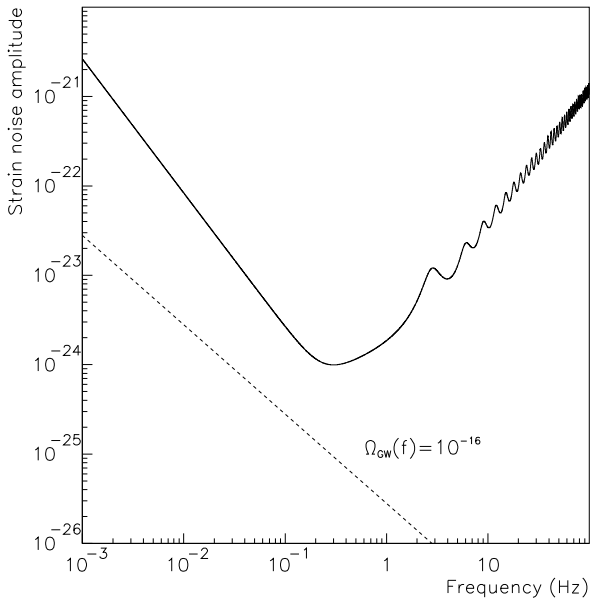}
  \caption{Amplitude of the BBO instrumental noise compared to primordial GW background.}
  \label{fig3}
 \end{figure}
In case the temperature around old neutron stars is  smaller or the disk heating 
efficiency is quite low, sub-millimeter emission detection would be privileged. 

Phillips and Chandler \cite{phch} reported  a millimeter and sub-millimeter disk search from five neutron stars:
two middle aged, isolated pulsars with characteristic ages of 10$^6$-10$^7$ years and three millisecond pulsars 
with ages of 10$^8$-10$^9$ years. The distance of these pulsars range between 100 pc and 3.6 kpc. No emission was found 
down to very low noise limits. Upper limits to disk masses of 10$^{-3}$-10$^{-4}$  M$_{\odot}$
for nearby pulsars and between  10$^{-3}$ and  a few solar masses for the most distant pulsars were 
consequently set.

%Kock-Miramond et al. \cite{kock} searched for evidence of emission at 15 $\mu$m from six nearby 
%pulsars both isolated and in binary systems  up to a distance of
%1 kpc. No emission were detected. For the nearest pulsar J0108-1431 the 3$\sigma$ upper limits on the flux density was 
%about 66 mJy at 15 $\mu$m and 22.5 mJy at 90  $\mu$m.  The authors set upper limits to the masses of circumpulsar dust 
%up to 1 kpc. 
More recently, Posselt et al. \cite{pose} presented the results of sub-millimeter measurements from 
RX J1856.5-3754, a neutron star located
at 167 pc. Observations could have been explained by a passively irradiated disk surrounding the star.
Under these conditions, a mass accretion 
of 10$^{14}$ g s$^{-1}$, 
a disk  inner radius of 10$^{14}$ cm and an upper limit to the disk  mass of a few Earth masses were estimated.

%Search for fallback disks are carried out in near-, mid-, far-infrared and optical observations. X-ray 
%emission is found to provide provide precious clues about the origin of the infrared observations from the disk.
In the case of the disk observed around the AXP 4U 0142+61, Wang, Kaplan and Chackrabarty \cite{wang} inferred from the 
comparison of X-ray, optical and mid-infrared observations that the disk was passively illuminated 
by the X-ray emission from the star. In addition, they found  that accretion was not the origin of the X-ray emission 
since accretion  would have generated an optical flux well above observations.
It is generally assumed that the X-ray emission is due to magnetar activity.
%More in general, the X-ray emission from neutron stars surrounded by disks depends from the characteristics
% of the interaction
%between stars and disks.  
In conclusion, at present time 
electromagnetic surveys  allow for the detection of disks with masses larger than 10$^{27}$ kg within a few kpc
distance, at most.

\subsection{Gravitational wave detection from circumpulsar precessing disks}

A  discussion on the possibility  to detect circumpulsar planetary systems and precessing disks
through gravitational wave observations near Earth
was presented in  \cite{grim09}.
The precession of an axisymmetric disk  occurs when the disk axis of rotation and symmetry axis are not aligned.
% Rigid-body disk precession around magnetized neutron stars was considered by Pfeiffer and Lai \cite{pfe}, for instance.
The frequencies
of gravitational waves generated by circumpulsar disk precession are $\omega$ and 2$\omega$, where $\omega$ is defined  as it follows
 \cite{grim09,lees}:

\begin{equation}
 \omega= \frac{I_3}{I_1 cos\theta} \Omega_3.
\end{equation}

We call $I_1$,  $I_2$  and $I_3$  the principal  moments of inertia with respect to
the principal axes,  $x_1$, $x_2$ and $x_3$, fixed in the disk. $\Omega_3$ is the angular velocity along the symmetry axis $x_3$   and $\theta$ is
the wobble angle defined as the angle between the angular momentum and the symmetry axis of the disk.
% Rigid-body disk precession around magnetized neutron stars was considered by Pfeiffer and Lai for instance \cite{pfe}.

 In the literature $\Omega_3$ is commonly assumed equal to the Keplerian frequency $\Omega_K$
\cite{lees} at the inner disk radius (see for example \cite{grim09,hake}).
We recall that the Keplerian frequency for the disk  is determined
 from the equation  $V$=$\Omega_K$$R$, where $V$ is the disk velocity at the inner radius,  
$R$, 
and $V$= $\sqrt{(GM_{o)}/R}$ where G and $M_{o}$ were defined in section 4.

Under the small wobble angle approximation ($\theta$=0.1 degrees),
%, if $\Omega_3$ ranges from the keplerian frequency associated with the
%disk dimensions and the
%frequency corresponding to the pulsar rotation,
the corresponding gravitational wave frequencies are  approximately 10.4(20.7)
Hz  for young pulsars and 0.23 (0.46) Hz
for middle aged pulsars.

The gravitational
wave amplitudes  (h) depend on the inclination angle ($i$) of the disk angular momentum with respect to the line
of sight (see \cite{lees} and references therein for details). For instance, in the case $i$$\simeq0$ and $\theta$ small:

\begin{equation}
 h \simeq  \frac{G}{ c^4} \frac{\omega^2}{r} \Delta I  \theta^2
\end{equation}
where r is the distance between the precessing disk and the observer and $\Delta I$ is defined as follows:

\begin{equation}
 I_1=I_2=\frac{1}{2} I_3=\Delta I.
\end{equation}

\begin{figure*}[t]
\includegraphics[width=\textwidth]{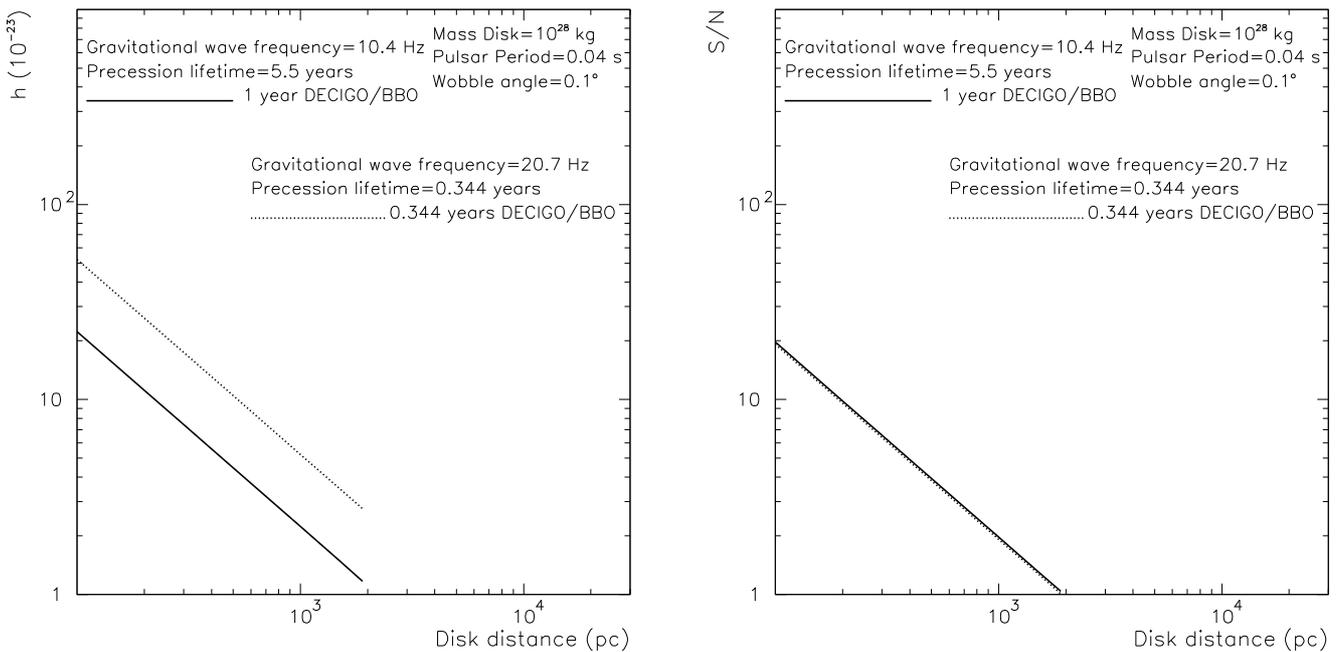}
\caption{\label{phe} Left panel. Amplitude of gravitational waves generated by Keplerian disks of 10$^{28}$ kg mass precessing around the light cylinder of young pulsars. Right panel. S/N ratio estimated for DECIGO/BBO.}
 \label{fig4}
\end{figure*}

If we take into account the component of the  radiation reaction torque
 perpendicular to the angular momentum, under the small angle
approximation only, the wobble angle change can be expressed as a function of $\theta$:

\begin{equation}
\dot{\theta}= -\frac{1}{\tau_\theta} \theta
\end{equation}

where:

\begin{equation}
\frac{1}{\tau_\theta} = \frac{2G}{5c^5} \frac{(\Delta I)^2} {I_1} \omega^4.
\end{equation}

For the  detection of gravitational waves generated by precessing circumpulsar disks  we
consider  the DECIGO
\cite{decigo1} and BBO \cite{bbo}  missions. 
%The DECIGO/BBO design were proposed as future space missions designed to operate in the range 0.1-1 Hz.
The DECIGO/BBO interferometers consist of a constellation of four triangular LISA (Laser Interferometer
Space Antenna)-like apparata \cite{lisaor} orbiting the Sun at 1 AU. In figure 2 we have reported 
the  BBO sensitivity curve \cite{bbo,jan3}. An analogous performance was predicted for DECIGO. 

%The  expected sensitivity of these experiments
%are of the order of  $10^{-24}$  between 0.5 and 1 Hz.
The BBO sensitivity curve  was estimated from the  experiment spectral
noise densities ($S^n(A)$; $S^n(E)$;$\ \ $ $S^n(T)$) reported in equations 8 and 9 as a function of frequency, f,  for the experiment uncorrelated channels A, E and T \cite{jan3,jan2}:

%\begin{equation}
%\begin{split}  
%S^n(A)= S^n(E)&=16\ sin^2(2\pi f L/c)\  (3+2\ cos(2\pi f L/c)\ +\ cos(4\pi f L/c))\ S^{tm}\ \\
%&+\ 8sin^2(2\pi f L/c) (2+cos(2\pi f L/c))\ S^{shot}\ \\ 
%S^n(T)&= 128\ sin^2(2\pi f L/c)\ sin^4(\pi f L/c)\ S^{tm}\ +\ 16(1-cos(2\pi f L/c)) \\
%&\ sin^2(2\pi f L/c)\ S^{shot}\ 
%\label{eq} 
%\end{split} 
%\end{equation}    

\begin{eqnarray}
\nonumber S^n(A)&=S^n(E)=16\ sin^2(2\pi f L/c)\  (3+2\ \\ \nonumber &cos(2\pi f L/c)\ +\ 
cos(4\pi f L/c))\ S^{tm}+\ \\ \nonumber &8\ sin^2(2\pi f L/c)\\ &(2+cos(2\pi f L/c))\ S^{shot}\ \\
\nonumber S^n(T)&= 128\ sin^2(2\pi f L/c)\ sin^4(\pi f L/c)\ S^{tm}\ +\\ \nonumber &16\ (1-cos(2\pi f L/c)) \nonumber \\
&\ sin^2(2\pi f L/c)\ S^{shot}\
\label{eq}
\end{eqnarray}
with test-mass noise $S^{tm}$=$S^{acc}$/(2$\pi$fc)$^2$  and shot noise $S^{shot}$=(h$\omega_o$/($P_{rec}$2$\pi$)) (2$\pi$f/$\omega_o$)$^2$
in terms of double sided spectral densities. 
We recall that the channels A, E and T are defined in terms of basic vectors X$_1$, X$_2$ and X$_3$. Each time-delay interferometry 
 X$_i$ mimic an unequal arm Michelson interferometer centered at the interferometer i. The standard design 
of  BBO provides a spectral density of test-mass acceleration $S^{acc}$=9$\times$10$^{-34}$ \\ m$^2$/s$^4$/Hz assumed
equal for all test masses and a light power, $P_{rec}$, equal to 9 W as  received by a spacecraft from one of its neighbors.
The carrier frequency of the laser is $\omega_o$=5.31$\times$10$^{15}$ s$^{-1}$and L=50000 km is the nominal arm 
length of the interferometer.
%  while decrease by more
%than 1.5 orders of magnitude at 10-20 Hz \cite{bbo} and more above these frequencies.

The signal-to-noise (S/N) ratio for a $quasi$ $periodic$ signal, can be estimated on the basis of  
the well approximated expression (see for example \cite{vetra}): 

\begin{equation}                                             
S/N \approx \frac{h_o}{S_n} \sqrt{\Delta t}                
\label{eq}                                                                                                                  
\end{equation}    

where $h_o$ is the amplitude of the GW signal, $S_n$ is the mean value of the linear power spectral density 
and $\Delta t$ is the observation time.

%In the  following, we estimate the  amplitudes  of the gravitational waves generated by precessing  disks around young and middle aged pulsars. 

\begin{figure*}[ht]
\includegraphics[width=\textwidth]{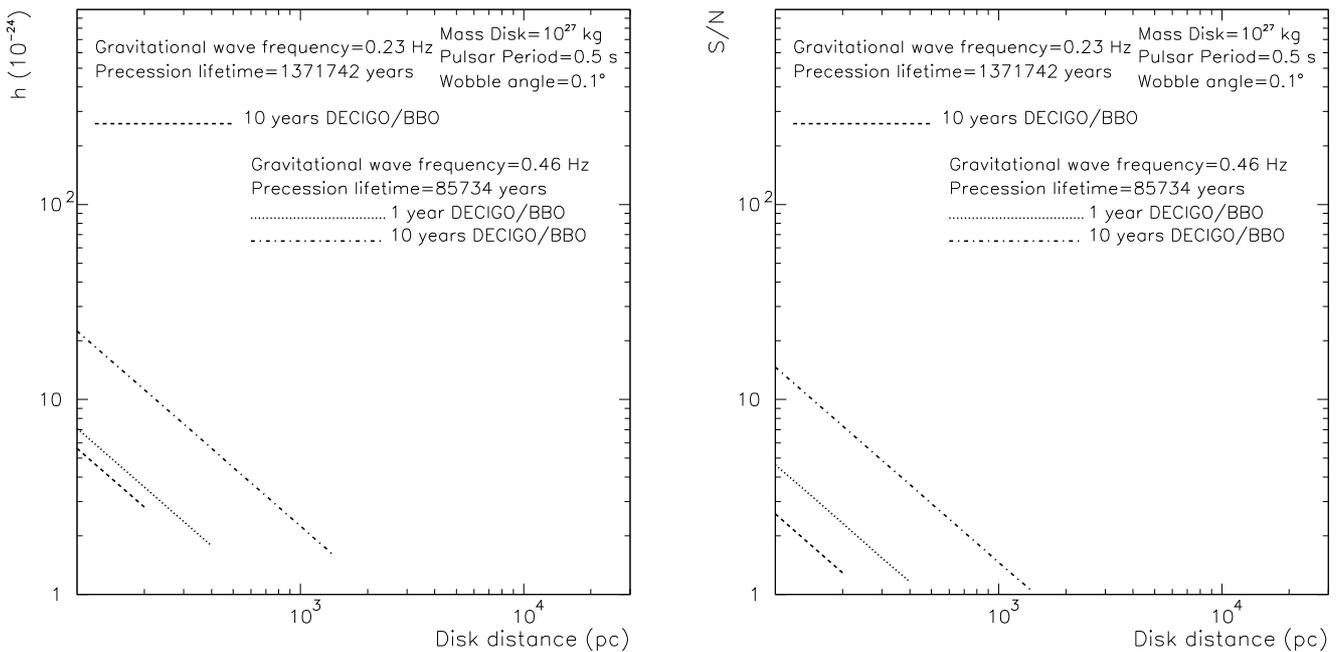}
\caption{\label{phe} Left panel. Amplitude of gravitational waves generated  by precessing Keplerian disks of 10$^{27}$ kg mass formed around the light cylinder of  middle aged pulsars. Right panel. S/N ratio estimated for DECIGO/BBO.}
 \label{fig5}
\end{figure*}

\begin{figure*}[ht]
\centering \includegraphics[width=\textwidth]{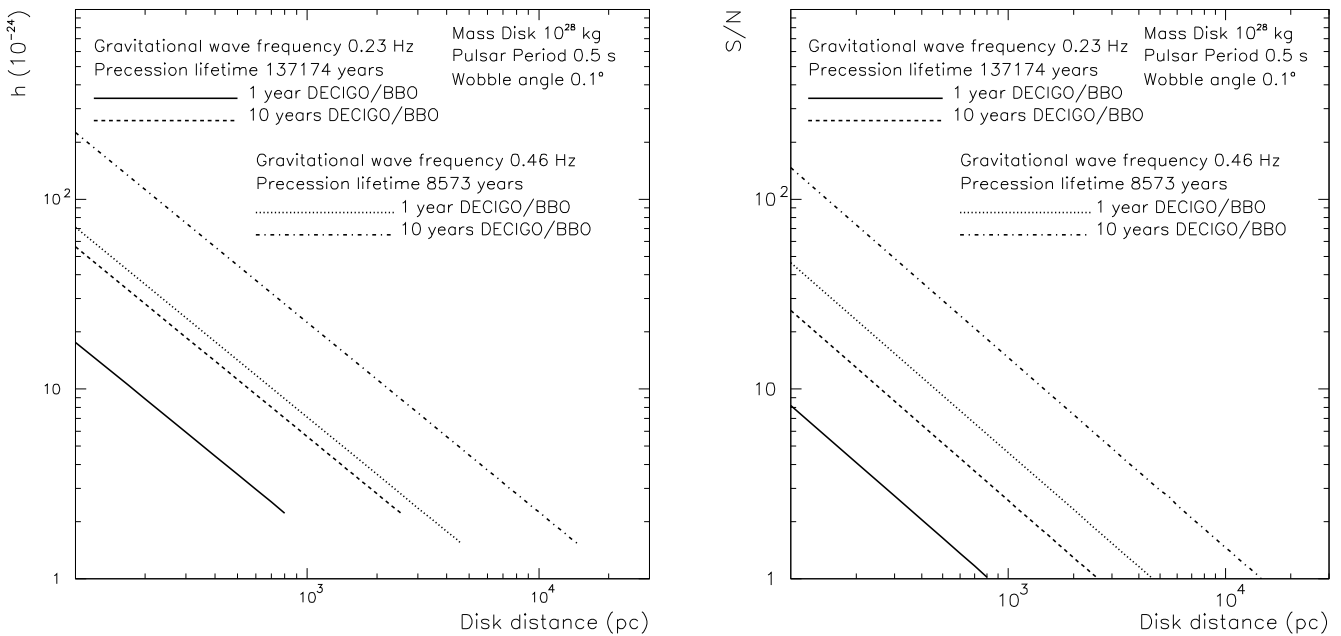}
\caption{\label{phe} Same as figure \ref{fig5} for  disk masses of 10$^{28}$ kg. Right panel. Same as figure 
\ref{fig5}.}
 \label{fig6} 
\end{figure*}

In the following we focus  on GWs generated by the precession of disks formed near the light cylinder of
young (hereafter pulsars with an age $<$ 10$^5$ years) and middle aged ( pulsars with an age $\le$
5$\times$10$^5$ years) pulsars since
%precessing circumpulsar disks formed near the light cylinder of middle aged pulsars generate 
these GWs lie in the
range of frequency of maximum sensitivity of DECIGO and BBO.
The S/N ratios for these gravitational waves possibly observed on DECIGO /BBO interferometers are also determined.
  We consider here wobble angles of  0.1 degrees and 5 degrees \cite{mont} 
%as parameters for $\theta$
 and disk masses ranging between 10$^{25}$ and 10$^{28}$ kg. DECIGO/BBO mission
possible durations of 1 and 10 years are assumed. Results are reported in figure \ref{fig4} 
for young pulsars and figures \ref{fig5} and \ref{fig6} for middle aged pulsars when  wobble angles of 0.1$^o$ are considered.
GW amplitudes appear in the left panels while S/N ratios appear in the right panels of these figures.
Gravitational waves of frequencies near 1 Hz
 are privileged for detection in terms of both gravitational wave amplitude
and disk precession lifetime.
Precessing disks around young and middle aged pulsars with masses larger than 10$^{27}$ kg 
and wobble angles of 0.1 degrees can be detected with S/N ratios larger than one within
a few kpc from the Solar System during one year  of  DECIGO/BBO mission duration,  while  heavier disks
 surrounding middle aged pulsars can be observed up to 20 kpc  for  ten year DECIGO/BBO mission duration.
% Gravitational waves of frequencies near 1 Hz
% are privileged for detection in terms of both gravitational wave amplitude
%and disk precession lifetime.
Disks around middle aged pulsars with masses above $10^{25}$ kg
would be detected for the whole disk precessing period  for  wobble angles of  5 degrees at 0.2-0.4 Hz
at distances larger than 10 kpc.
%In figure \ref{fig6} it can be noticed that disks surrounding young pulsars can be observed at 10-20 Hz within 1 kpc
%from the solar system for large disk masses only.
%However, for increasing disk masses the precessing time decreases down to a few tens of years.
Detection appears extremely unlikely  for gravitational wave frequencies above 30 Hz due to small
disk precessing time.

 We point out that gravitational wave signals  generated by precessing disks  below 1 Hz lie  above 
those expected from
primordial gravitational waves as it can be observed in figure \ref{fig3} \cite{bbo}.
Coalescing  intermediate-mass black hole binaries could also be  feasible sources of gravitational waves
in the same frequency range of circumpulsar precessing disks \cite{decigo1}.
However,  doubts remain about
the very existence of intermediate-mass black holes \cite{mapelli}.

The realization of future space interferometers appears more feasible after the European Space Agency in November 2013
selected the $Gravitational$ $Universe$ as one of its corner-stone science themes. In particular, the eLISA
mission should be the first interferometer devoted to gravitational wave detection in space operating in the
interval 10$^{-4}$-10$^{-1}$ Hz. The maximum sensitivity expected for eLISA is 3$\times$10$^{-21}$ at 10$^{-2}$ Hz
\cite{elisacon},
while at 1 Hz is six orders of magnitude smaller than that expected for DECIGO/BBO.
eLISA would be able to  detect precessing circumpulsar
disks within 100 pc between 10$^{-2}$ Hz and 1 Hz with a S/N ratio of 1 with 10 year mission by assuming
wobble angles of at least a few degrees. 
Unfortunately, as we will show in the next section, no more than one middle aged pulsar-disk system
is expected to be found within this distance.

Therefore,  eLISA is not expected to be able to detect circumpulsar precessing disks.
%In other words, we don't expect eLISA
%to detect the presence of circumpulsar precessing disks.

\section{Upper limit to the detection rate of circumpulsar disks with DECIGO/BBO}

In order to estimate the possible  number of detectable pulsar-disk systems in the Sun environment,
we use  the results of the work reported by Jiang and Li in \cite{jili}. These authors simulated
 the evolution of 2$\times$10$^6$ neutron stars up to an age of 10$^4$ years
 by assuming a possible disk formation in the whole sample of the studied stars.
Plausible values following from experimental evidences and theoretical models for neutron star surface
 magnetic fields, initial periods and disk masses were
used in the simulations. For the star surface magnetic fields, a log-normal distribution of mean 12.5 and standard
deviation 0.3 was considered. The neutron star initial periods were assumed uniformly distributed
between 10 ms and  100 ms and for the initial mass of the disks the log($\dot{M_o}$ $t_o$/M$_{\odot}$)
was uniformly distributed between -6 and -2. It was found that depending on the propeller
mechanism actually applying, a fraction ranging between 25\% and 50\% of pulsars is
compatible with surrounding disks in the age range 10$^3$- 10$^5$ years. 
%As it was recalled above, disks older than
%10$^5$ years are supposed to be neutral \cite{cordes}.
%On the basis of the work by  Ek\c{s}i, Hernquist and Narayan  \cite{eksi}, no disk can form
%for pulsar initial periods smaller than approximately 40 ms, 
We conclude that, on the basis of the results reported in section 3, the upper limit of 50\% of pulsars
surrounded by disks follows from considering  initial periods
 smaller than 50 ms for half of the simulated pulsar sample. In conclusion,
 this simulation work indicates that, in principle, the whole sample of
isolated pulsars could be surrounded by disks.  
%unless pulsar initial periods are, on average, too small to allow for disk formation.

\begin{table}
\caption{\label{table1} Estimates of the number  of active pulsars in the Milky Way.
Table reference legend: VN81 \cite{vn81}; LMT85 \cite{l85}; N87 \cite{n87} ; LEA93 \cite{le93}; VEA04 \cite{vea04}. }
%\begin{center}
%\lineup
\begin{tabular}{lll}%{@{}*{2}{l}}
\hline\noalign{\smallskip}
Author & Number of active pulsar  \\
\noalign{\smallskip}\hline\noalign{\smallskip}
VN81       & 6$\times$10$^5$      \\
LMT85        & 2$\times$10$^5$      \\
N87       &  1.5$\times$10$^5$    \\
LEA93       &  7$\times$10$^4$    \\
VEA04       &  7$\times$10$^4$-1.2$\times$10$^5$    \\
\noalign{\smallskip}\hline\noalign{\smallskip}
\end{tabular}
%\end{center}
\end{table}
%As a following step, we consider
%to estimate the number of possible detections of precessing disks around young 
%pulsars (up to 10$^{5}$ years), we have taken into account 
The spatial distribution of pulsars
in the Galaxy can be inferred from \cite{faucher,guse}. 
%\subsection{Pulsar spatial distribution}
%The average number of pulsars  in the solar
%environment can be estimated on the basis of the  PB and the pulsar
%distribution in the galactic disk.
%The pulsar spatial distribution was studied, for instance, in
%\cite{faucher}  and  \cite{guse}.
In these works the authors show that pulsars are concentrated in the Milky Way spiral arms
and present  a radial distribution with  a maximum at 3.5 kpc from the galactic center (GC).
%Similar results are obtained by  Guseinov and Kosumov
%\cite{guse}.
On the basis of figure \ref{fig3}  in both  above papers,
we find   that approximately 13\% of pulsars can be found in the region of the Galaxy between 7.5 and 9.5 kpc
from the GC.
This fraction increases to 36\%  between 6 and 11 kpc from the GC.
We recall that the Solar System lies at 8.5 kpc from the GC.

By assuming a PB of one pulsar born every 60 years (see section 3)
and an upper limit to the active
 pulsar lifetime of   2$\times$10$^7$ years \cite{michwo},
the maximum number of active pulsars in our galactic disk comes to be  3.3$\times$10$^5$. This simple estimate is consistent
 with the results obtained with radio surveys as it can be observed in
Table 2.
%In this work we focus  on young (hereafter pulsar with an age up to 10$^5$ years) and middle aged ( pulsar with an age up to 
%5$\times$10$^5$ years) pulsars since
%precessing circumpulsar disks formed near the light cylinder of middle aged pulsars generate gravitational waves in the 
%range of frequency of maximum sensitivity of DECIGO and BBO.
%older than 10$^{5}$ years are  considered to be passive \cite{cann}. 
%We recall that
% middle aged pulsars  were also indicated as the main sources of positrons in
%cosmic rays above a few tens of GeV (\cite{gri07aa} and references therein).
A uniform distribution of active pulsars  versus age indicates that  a fraction of 2.5\% of the galactic
 sample is constituted by pulsars  younger than
5$\times$10$^5$ years: this amounts  to  8250 pulsars.

The average number density of young and middle aged pulsars within 1 kpc from the Solar System is, for instance, 33 pulsars 
kpc$^{-3}$.
This result follows from the previously mentioned figure 2 of the works by Faucher-Gigu\`ere and Kaspi \cite{faucher} and Guseinov and Kosumov  \cite{guse}
 by considering an effective
Milky Way disk volume of 32.8 kpc$^{3}$ for the region between 7.5 and 9.5 kpc from the GC.\\
\begin{figure}[t]
  \vspace{5mm}
  \centering
  \includegraphics[width=3.5in]{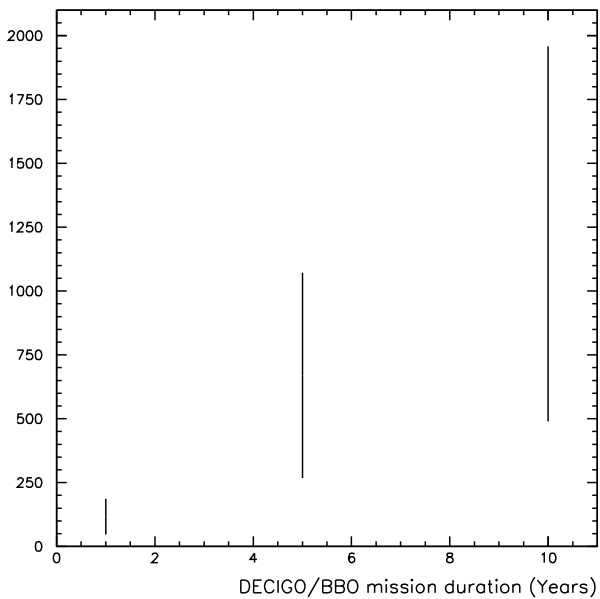}
  \caption{Estimate of the minimum and maximum numbers of precessing  disk detections around middle aged pulsars versus DECIGO/BBO mission duration. Disks with masses larger than 10$^{27}$ kg  and  wobble angles of 0.1 degrees were considered.}
  \label{fig7}
 \end{figure}
%For the results reported in the following we assume that 
%will call young pulsars those with ages smaller than 10$^5$ years and that 
%disk masses are uniformly distributed between 10$^{25}$ kg and $^{28}$ kg. 
\indent In the case of young pulsars and small wobble angles, from figure  \ref{fig4}
we find that only those stars within 1 kpc with disk masses larger than 5$\times$10$^{27}$ kg
could be detected. Young pulsars have a number density of 11 kpc$^{-3}$. 
By  assuming that disk masses are uniformly distributed between 10$^{25}$ kg and $^{28}$ kg,
 a fraction of 2.5$^{-1}$
of disks could be detected because of the minimum required mass.
The minimum and maximum numbers of detections are set on the basis of the work by \cite{jili}.
We find a maximum number of possible detections of 4 at 10.4 Hz, the minimum number would be 1 
by assuming one year DECIGO/BBO mission.
% for disk masses larger than
%5$\times$10$^{27}$ kg. 
For middle aged pulsars (10$^{5}$-5$\times$10$^{5}$ years) and 
small wobble angles, the higher detection rate is expected at 0.46 Hz for disk masses larger than 
10$^{27}$ kg (see figure \ref{fig6}). 
%Minimum and maximum values are set following the 
%results of the simulation by Jiang and Li \cite{jili}. 
Results are shown in figure \ref{fig7}.

In case of wobble angles of a few degrees, up to  thousands of precessing disks  could be detected 
 up to tens of kpc distance in one year of DECIGO/BBO mission. 
%In subsection 3.2 we have found that the average number density of
%young and middle aged pulsars within one kpc from the Sun is 33 kpc$^{-3}$. The volume of the Galaxy
%actually hosting pulsars within one kpc
%from the Sun  is 0.96 kpc$^{-3}$, therefore the  number of detectable precessing circumpulsar disks
%within one kpc from the solar system ranges between 8 and 32 
%during the minimum expected duration of space interferometer missions of one year
%following the work by Jiang and Li.
%Cosmic-ray positrons observed near Earth at energies larger than  tens of GeV  are produced within 
% one kpc from the Solar System (see for example \cite{adri11}). In case  they originate in the magnetosphere of middle 
%aged pulsars, the characteristics of circumpulsar disks inferred from gravitational wave measurements should be properly 
%taken into account in $e^+$ production models.

%\begin{figure}[!t]
%  \vspace{5mm}
%  \centering
%  \includegraphics[width=3.5in]{fig3_app.eps}
%  \caption{Amplitude of gravitational waves generated  by precessing keplerian disks of 10$^{27}$ kg formed around the light cylinder of  middle aged pulsars.}
%  \label{fig4}
% \end{figure}

%\begin{figure}[!t]
%  \vspace{5mm}
%  \centering
%  \includegraphics[width=3.5in]{fig2_app.eps}
%  \caption{Same as figure \ref{fig4} for  disk masses of 10$^{28}$ kg.}
%  \label{fig5}
% \end{figure}

%\begin{figure}[!t]
%  \vspace{5mm}
%  \centering
%  \includegraphics[width=3.5in]{fig4_app.eps}
%  \caption{Amplitude of gravitational waves generated  by  disks of 10$^{28}$ kg precessing around young pulsars.}
%  \label{fig6}
% \end{figure}

\section{Conclusions}

Pulsar average initial parameters
are compatible with the possibility that
a large fraction of young and middle
aged pulsars are  surrounded by  disks. Precessing Keplerian disks near the light cylinder of 
young and middle aged
pulsars are expected to generate gravitational waves in the frequency range  0.2-20 Hz. Detection of these 
gravitational waves  appears feasible near one Hz
with the  future DECIGO/BBO space interferometers.
% during  one year mission duration for disk distances smaller than one kpc.
%This detection could be extended up 
Up to thousands of disk-pulsar systems could  conceivably be detected up to
 tens of kpc distance during ten year space interferometer  mission durations and disk masses larger
than 10$^{27}$ kg under the  small wobble angle approximation. Gravitational waves generated by smaller mass 
precessing  disks  could be observed for  wobble angles of a few degrees.

The observations of gravitational waves generated by circumpulsar disks will
allow us to estimate the  disk geometrical characteristics and to
study their role in quenching
the particle production in the  pulsar magnetosphere.

Cosmic-ray positrons observed near Earth at energies larger than  tens of GeV  are produced within
 one kpc from the Solar System. In case  they originate in the magnetosphere of middle
aged pulsars, the characteristics of circumpulsar disks inferred from gravitational wave measurements should be properly
taken into account in $e^+$ production models.

\subsection*{\bf Acknowledgments}

This research work was funded by the Department of Basic Sciences 
and Fundamentals of the University of Urbino ``Carlo Bo'', Italy.
The author is very grateful to Dr. J. Harms of the INFN-Florence (Italy) for providing precious comments and data to reproduce
figure 3. A special thank is due to K. J. Kieswetter for kindly proofreading the manuscript and to M. Fabi for technical support.

\footnotesize{}

%\begin{acknowledgements}
%\end{acknowledgements}



\end{document}